\documentstyle[epsf,twocolumn,prl,aps]{revtex}  
\draft
\title{A unifying model for chemical and colloidal gels 
}
\author{Emanuela Del Gado,$^{a,c}$, Annalisa Fierro,$^{a,c}$, 
Lucilla de Arcangelis, $^{b,c}$ and Antonio Coniglio $^{a,c}$}
\address{${}^a$Dipartimento di Scienze Fisiche, Universit\`a di Napoli
         "Federico II",\\ Complesso Universitario di Monte Sant'Angelo,
         via Cintia 80126 Napoli, Italy}
\address{${}^b$Dipartimento di Ingegneria dell'Informazione, 
         Seconda Universit\`a di Napoli, via Roma 29,
 81031 Aversa (Caserta), 
         Italy}
\address{${}^c$ INFM Udr di Napoli and Gruppo coordinato SUN}
\date{April 4, 2003}
\begin{document}
\maketitle
\begin{abstract}

We investigate the slow dynamics in gelling systems by means of 
MonteCarlo simulations on the cubic lattice of a minimal statistical mechanics 
model. By opportunely varying some model parameter 
we are able to describe a crossover from the chemical gelation behaviour to 
dynamics more typical of colloidal systems. 
The results suggest a novel connection linking 
classical gelation, as originally described by Flory, 
to more recent results on colloidal systems.
PACS: 05.20.-y, 82.70.Gg, 83.10.Nn
\end{abstract}
\begin{figure}
\begin{center}
\mbox{ \epsfxsize=8cm
       \epsfysize=8cm
       \epsffile{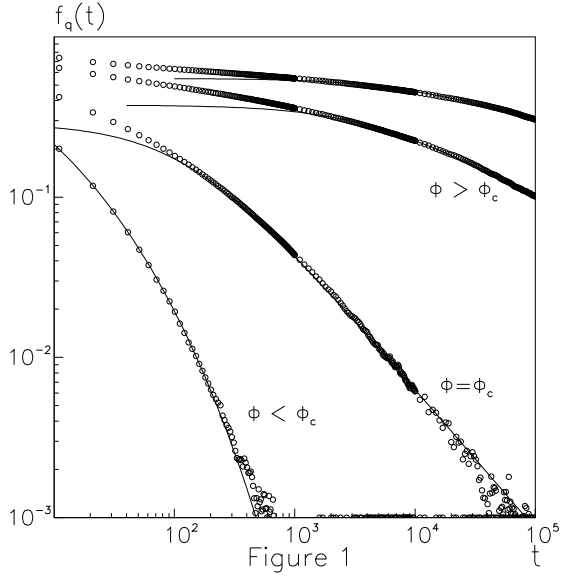}
      }
\end{center}
\caption{Double logarithmic plot of the autocorrelation functions 
$f_{\vec{q}}(t)$ as function of the time for $q \sim 1.36$ and 
$\phi=0.6,0.718,0.8,0.85$. 
For $\phi < \phi_{c}$ the long time decay is well fitted by a  
function (full line) $\sim e^{-(t/\tau)^{\beta}}$ with $\beta \sim 0.3$. 
At the percolation threshold and in the gel phase in the long time decay the 
data are well fitted by a function $\sim (1 + \frac{t}{\tau'})^{-c}$. 
}
\label{figure1}
\end{figure}
\begin{figure}
\begin{center}
\mbox{ \epsfxsize=7.5cm
       \epsfysize=7.5cm
       \epsffile{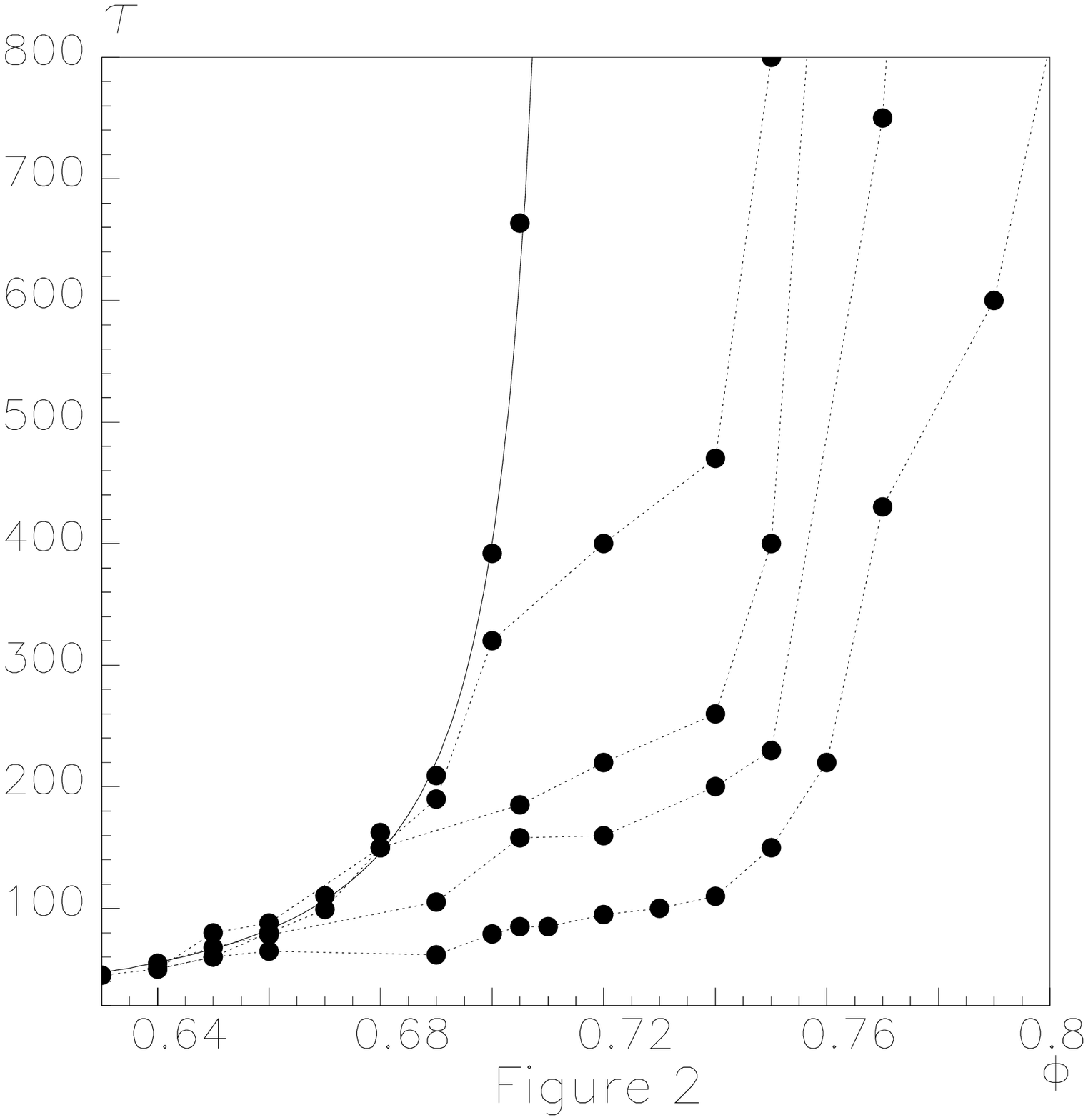}
      }
\end{center}
\caption{
The average relaxation time as function of the density; from left to right:
the data for the permanent bonds case 
diverge at the percolation threshold with a power law (the full line); 
the other data refer to finite $\tau_{b}=3000,1000,400,100 MC step/particle$ 
decreasing from 
left to right (the dotted lines are a guide to the eye) 
}
\label{figure2}
\end{figure}
\begin{figure}
\begin{center}
\mbox{ \epsfxsize=8.6cm
       \epsfysize=8.6cm
       \epsffile{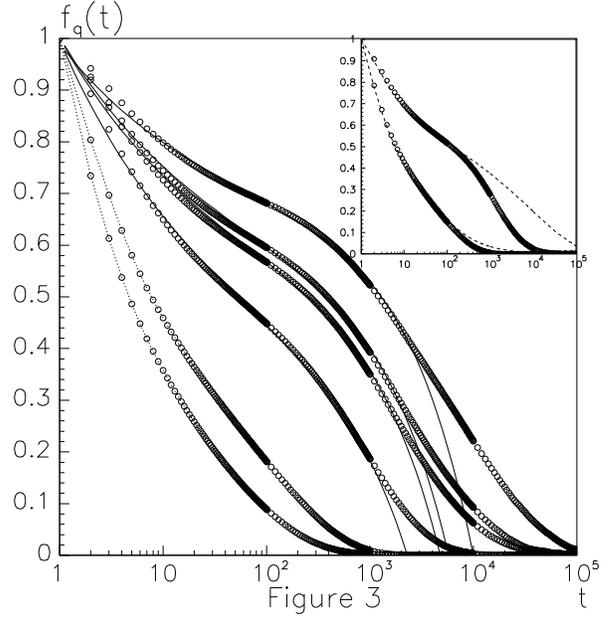}
      }
\end{center}
\caption{
The 
$f_{\vec{q}}(t)$ as function of the time for $q \sim 1.36$
calculated on a cubic lattice of size $L=16$ for 
$\tau_{b}=1000 MCstep/particle$. (from left to right $\phi= 0.6,0.718,0.8,
0.85,0.87,0.9$). The full lines correspond to the fit with
the mode-coupling $beta$-correlator. In the inset,  
the data for $\phi= 0.718$ and $\phi=0.8$ (from left to right) 
are plotted together with 
the ones obtained at the same densities in the case of permanent bonds 
(dashed line).
}
\label{figure3}
\end{figure}
\begin{figure}
\begin{center}
\mbox{ \epsfxsize=7.5cm
       \epsfysize=7.5cm
       \epsffile{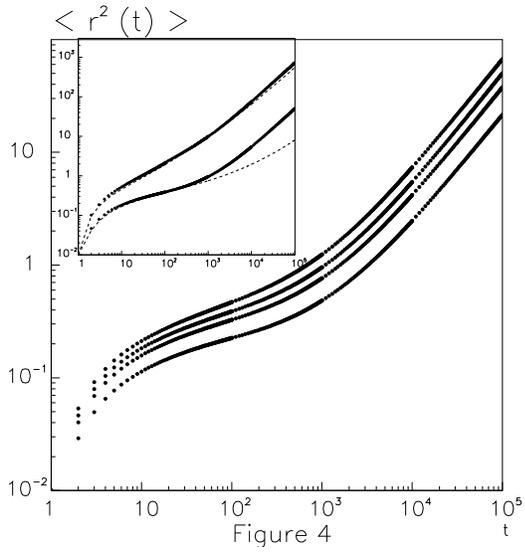}
      }
\end{center}
\caption{
The mean-square displacement $\langle \vec{r}^{2}(t) \rangle$ of the 
particles as function of the time in a double logarithmic plot for 
$\tau_{b} = 1000 MCstep/particle$: from top to 
bottom $\phi=0.8,0.82,0.85,0.9$, approaching $\phi_{g}$.  
In the inset $\langle \vec{r}^{2}(t) \rangle$ for the same $\phi$  
for $\tau_{b}=1000 MCS/particle$ and in the case of permanent bonds 
(dashed line) (from top to bottom, $\phi= 0.718,0.8$)
}
\label{figure4}
\end{figure}

%
%
%%%%%%%%%%%%%%%%%%%%%%%%%%%%%%%%%%%%%%%%%%%%%%%%%%%%%%%%%%%%%%%%%%%%%
%
\vspace*{0.8cm}
The chemical gelation transition, as it is typically observed in polymer 
systems, transforms a solution of polymeric molecules, the sol, from a 
viscous fluid to 
an elastic disordered solid, the gel. 
The viscosity coefficient grows with a power law behaviour, as function of the
relative difference from the critical polymer concentration, 
characterized by a 
critical exponent $k$. The onset of the elastic response in the system, 
as function of the same control parameter, 
displays a power law increasing of the elastic modulus described
by a critical exponent $f$ \cite{viscoela}. This corresponds to the
constitution inside the sol of a macroscopic polymeric structure \cite{flory},
that characterizes the gel phase. As implicitly suggested in the
work of Flory and Stockmayer \cite{flory}, the percolation model
is considered as the basic model for the chemical gelation transition and the 
macromolecular stress-bearing structure in these systems is a percolating 
network \cite{degl,stau,adconst}.\\ 
Moreover slow dynamics characterize the gelling solution: 
the relaxation functions in the experiments display a long time stretched 
exponential decay $\sim e^{-(\frac{t}{\tau_0})^{\beta}}$ 
as the gelation threshold is approached. 
The relaxation process becomes critically slow at the gel point, where the 
onset of a power law decay is typically observed \cite{relax}.\\
Gelation phenomena are also observed in colloidal systems, that are
suspensions of mesoscopic particles interacting via short range attraction: 
due to aggregation phenomena at low density (colloidal gelation)
these systems display gel states with 
a power law behaviour of the viscosity coefficient and of the elastic 
modulus \cite{weitzna}, as in chemical gelation. Yet the 
aggregation process gives rise to cluster-cluster aggregation 
producing a spanning cluster with a fractal dimensionality 
smaller than the random percolation case \cite{meakin,jullien,gimel,dinsweitz}. 
On the other hand with a weaker attraction at higher density a gelation 
characterized by a glass-like kinetic arrest \cite{weitzclu,relaxco} 
may be observed. The relaxation patterns closely recall the 
ones observed in glassy systems and are well fitted 
by the mode-coupling theory \cite{goe} predictions for supercooled 
liquids approaching the glass transition \cite{relaxco}. 
On the theoretical side 
the application of the mode-coupling theory to 
systems with short range attractive interaction \cite{mct,ema,cates} 
({\em attractive glasses}) has been recently considered and the connection with the colloidal 
glass transition has been proposed.\\ 
The question, that we want to investigate in this paper, is whether
and to what extent colloidal gelation, colloidal glass transition and 
chemical gelation are related and if a unifying description is possible.\\
Let us first consider the case of the chemical gelation. 
By means of MonteCarlo numerical simulations on the cubic lattice we study a 
solution of tetrafunctional monomers. Each monomer occupies a lattice 
elementary cell and, to take into account the excluded volume interaction, 
two occupied cells cannot have common sites. 
At $t=0$ we fix the fraction $\phi$ of present monomers respect to the maximum
number allowed on the lattice, and randomly 
quench bonds between them. This actually corresponds to the typical chemical 
gelation process that can be obtained by irradiating the monomeric solution. 
The four possible bonds per monomers, randomly selected, are formed with 
probability $p_{b}$ along lattice directions between monomers that are nearest 
neighbours and next nearest neighbours. Once formed, the bonds are permanent. 
The monomers diffuse on the lattice via 
random local movements and the bond length may vary but not 
be larger than $l_{0}$ according to bond-fluctuation dynamics \cite{carkr}. 
The value of $l_{0}$ is determined by the self-avoiding 
walk condition and on the cubic lattice is $l_{0}=\sqrt{10}$ in lattice 
spacing units.
We let the monomers diffuse to reach the 
stationary state and then study the system for different values of the 
monomers concentration. 
We have considered $p_{b}=1$, for which the 
system presents a percolation transition at $\phi_{c} = 0.718 \pm 0.005$. 
We have introduced this lattice model  
\cite{dedecon3,dedecon4} to study the viscosity by means of the 
relaxation time, that diverges 
as function of $\phi$ at the percolation threshold with a power law 
behaviour \cite{dedecon3}. 
The elastic response in the gel phase has been studied by means of the 
fluctuations in the free energy and grows with a power law behaviour as well
as function of $(\phi -\phi_{c})$ \cite{dedecon4}. In this letter instead 
we present a completely new study of the dynamics in gelation phemomena.\\ 
To investigate the nature of the dynamic transition at the chemical gelation 
here we study the equilibrium density fluctuation autocorrelation 
functions $f_{\vec{q}}(t)$ given by
\begin{equation}
f_{\vec{q}}(t) = \frac{< \rho_{\vec{q}} (t+t')
 \rho_{\vec{q}}(t')>}{<|\rho_{\vec{q}}(t')|>^{2}}
\label{autct}
\end{equation}
where $\rho _{\vec{q}}(t)= \sum_{i=1}^{N} e^{-i \vec{q} \cdot \vec{r}_{i}(t)}$,
$\vec{r}_{i}(t)$ is
the position of the $i-th$ monomer at time $t$, $N$ is the number of monomers
and the average $\langle ... \rangle$ is performed over the time $t'$. Due to
the periodic boundary conditions the values of the wave vector
$\vec{q}$ on the cubic lattice are $\vec{q}=\frac{2\pi}{L}(n_{x},n_{y},n_{z})$
with $n_{x},n_{y},n_{z}=1...L/2$ integer values.\\ 
We also study the mean square displacement of the particles
$\langle \vec{r}^{2}(t) \rangle = \frac{1}{N} \sum_{i=1}^{N} \langle 
(\vec{r}_{i}(t+t') - \vec{r}_{i}(t') )^{2} \rangle$.\\
In Fig.\ref{figure1} we present these time autocorrelation functions as 
function of the time calculated on a cubic lattice of size $L=16$. 
The data have been averaged over $\sim 10$ up to $10^5$ 
time intervals and over $20$ different initial configurations of the sample. 
As the monomer concentration $\phi$ approaches the percolation threshold 
$\phi_{c}$, $f_{\vec{q}}(t)$ displays a 
long time decay well fitted by a stretched exponential law 
$\sim e^{-(t/\tau)^{\beta}}$ with a 
$\beta \sim 0.30 \pm 0.05$. At the percolation threshold the onset of a power 
law decay is observed as it is shown by the double logarithmic plot 
of Fig.\ref{figure1} with an exponent $c$ \cite{relax}. 
As the monomer concentration is increased above the percolation threshold in 
the gel phase, the long time power law decay of the relaxation functions can 
be fitted with a decreasing exponent $c$, varying from $c\sim 1.$ at 
$\phi_{c}$ to $c\sim0.2$ well above $\phi_{c}$, where a nearly logarithmic 
decay appears.  
This behaviour well agrees with the one observed in gelling systems 
investigated in the experiments of refs.\cite{relax}. It is interesting to 
notice that this kind of decay with a stretched exponential and a power law 
reminds the relaxation behaviour found in spin-glasses \cite{ogi}.\\ 
The mean square displacement of the particles
$\langle \vec{r}^{2}(t) \rangle$ 
presents a long-time diffusive behaviour and the diffusion coefficient 
decreases but remains finite also above $\phi_{c}$. 
However the diffusion coefficient of clusters of size 
comparable with the connectedness length goes to zero at $\phi_{c}$ with the 
same exponent as the relaxation time \cite{dedecon3}.\\ 
In colloids the aggregation is due to a short range attraction and in general 
the monomers are not permanently bonded. To take into 
account this feature we introduce a novel ingredient in the previous model 
by considering a finite bond lifetime $\tau_{b}$ and study the effect on the 
dynamics. It is worth noticing that the features of this model with finite 
$\tau_{b}$ can be realized in 
a microscopic model: a solution of monomers interacting via an attraction of 
strength $-E$ and excluded volume repulsion. Due to monomers diffusion the 
aggregation process eventually takes place. 
The finite bond lifetime $\tau_{b}$ corresponds to an attractive interaction
of strength $-E$ that does not produce permanent bonding between monomers,
and $\tau_{b} \sim e^{E/KT}$. 
A more detailed discussion   
will be available in a longer paper following up \cite{new}.\\   
Due to the finite $\tau_{b}$, in the simulations during the monomer diffusion 
the bonds between monomers are broken with a frequency $1/\tau_{b}$. 
Between monomers separated by a distance less than $l_{0}$ 
a bond is formed with a frequency $f_{b}$. 
For each value of $\tau_{b}$ we fix $f_{b}$ so 
that the fraction of present bonds is always the same \cite{new}.\\ 
After 
the system has reached the equilibrium we have calculated $f_{\vec{q}}(t)$ 
as defined in eq.(\ref{autct}) and the relaxation time $\tau$ as 
$f_{\vec{q}}(\tau) \sim 0.1$.\\   
Let us first analyze the behaviour of $\tau$, plotted in 
Fig.\ref{figure2} as function of the 
monomer concentration $\phi$ for different $\tau_{b}$.
For comparison we have also shown the behaviour of $\tau$ in the case of 
permanent bonds, which displays a power law 
divergence at the percolation threshold $\phi_{c}$.
We notice that for finite bond lifetime $\tau_{b}$  
the relaxation time increases following the permanent bonds case 
(chemical gelation), up to some value $\phi^{*}$ and then deviates from it. 
The longer the bond lifetime the higher $\phi^{*}$ is. 
In the high monomer concentration region, 
well above the percolation threshold, the relaxation 
time in the finite bond lifetime case again displays a steep increase and a 
power law divergence at some higher value. 
This truncated critical behaviour followed by a glassy-like transition 
has been actually detected in some colloidal systems in the viscosity
behaviour \cite{malla,durand}.\\ 
These results can be explained by considering that
only clusters whose diffusion relaxation time is smaller than $\tau_{b}$
will behave as in the case of permanent bonds. Larger clusters will not
persist and their full size will not be relevant in the dynamics: the
finite bond lifetime induces an effective cluster size distribution with a
cut-off, which keeps the macroscopic viscosity finite \cite{conmess}. As
the concentration increases the final growth of the relaxation time is due
to the crowding of the particles.\\ In Fig.\ref{figure3} $f_{\vec{q}}(t)$
is plotted as function of time for a fixed value of $\tau_{b}$ for
increasing values of the monomer concentration ($\phi$ varies from below
to well above $\phi_{c} = 0.718$). For small concentrations the
autocorrelation function $f_{\vec{q}}(t)$ is well fitted by a stretched
exponential decay, while for high monomer concentrations it exhibits a
two-step decay, that  closely resembles the one observed in supercooled
liquids. We fit these curves using the mode-coupling $\beta$-correlator
\cite{goe}, corresponding to a short time power law $\sim f + \left(
\frac{t}{\tau_{s}} \right)^{-a}$ and a long time von Schweidler law $\sim
f - \left( \frac{t}{\tau_{l}} \right) ^{b}$, giving the exponents $a \sim
0.33 {\pm} 0.01$ and $b \sim 0.65 {\pm} 0.01$ (the full lines in
Fig.\ref{figure3}).
At long times the different curves obtained for
different $\phi$ collapse into a unique master curve by opportunely
rescaling the time via a factor $\tau(\phi)$. The master curve is well
fitted by a stretched exponential decay with $\beta \sim 0.50 {\pm} 0.06$. The
characteristic time $\tau(\phi)$ diverges at a value $\phi_{g} \sim 0.96
{\pm} 0.05$ with the exponent $\gamma \sim 2.3 {\pm} 0.1$. This value well
agrees with the mode-coupling prediction $\gamma = 1/2a + 1/2b$
\cite{goe}.\\ In the inset the data are compared with the case of
permanent bonds for two different values of $\phi$. We observe that for
short time (of the order of $\tau_{b}$) the autocorrelation functions
coincide with the permanent bond case.\\   
A similar pattern, characterized by a plateau, is found in the mean square
displacement $\langle \vec{r}^{2}(t)\rangle$
(Fig.\ref{figure4}),with the diffusion coefficient going to zero as $\phi$
approaches $\phi_{g}$ . The inset also indicates that for short time the
mean square displacement coincides with that calculated for the
permanent bond case.\\ These results show that when bonds are permanent
(chemical gelation) the divergence of the relaxation time is due to the
formation of a macroscopic critical cluster and the autocorrelation function
exhibits a one step decay related to the relaxation of such  cluster. In
the case of finite $\tau_{b}$ there is an effective cluster size distribution, 
as discussed above, and the autocorrelation function exhibits a two step 
relaxation which is well fitted by the mode coupling theory. This can be
explained if we consider the system made of effective clusters playing the
role of single molecules in an ordinary supercooled liquid or in a colloidal 
hard sphere system. As the monomer concentration increases, the relaxation time will diverge due to the jamming of the effective clusters. 
The first decay is due to the rattling and relaxation of the "effective"
cluster in a cage made of surrounding clusters. The second relaxation is
due, as usual, to the opening of the cage resulting in a structural
relaxation. This picture supports the jamming of clusters that has been 
suggested on the basis of experimental observations on colloidal gelation in 
ref. \cite{weitzclu}.
For high temperature $\tau_{b} \rightarrow 0$,the clusters
reduces to single monomers with a crossover to the usual
phenomenology of the hard spheres glass transition.\\
When $\tau_{b}$ is large enough (strong attraction) the cluster effect
will dominate and the slow dynamics will exhibit features more closely
related to chemical gelation (Fig.\ref{figure1}). The only difference is
that in the limit $\tau_{b} \rightarrow \infty$ we expect that the
spanning cluster will have the structure of the cluster-cluster
irreversible aggregation model instead of random percolation.
\\
In conclusion, these results suggest a unifying approach for chemical
gelation, colloidal gelation and colloidal glass transition. In chemical
gelation and colloidal gelation the cluster formation should produce the
slow dynamics, that is expected to be of the same type of Fig.
\ref{figure1}. In colloidal systems for weak attraction and high
concentration the system crosses over from colloidal gelation to colloidal
glass due to the jamming of effective clusters.\\
We would like to thank
K. Dawson, G. Foffi, W. Kob, F. Mallamace, N. Sator, F. Sciortino, 
P. Tartaglia and E. Zaccarelli for many interesting discussions.    
This work has been partially supported by MIUR-FIRB-2002 and
by the INFM Parallel Computing Initiative.
%
%%%%%%%%%%%%%%%%   REFERENCES  %%%%%%%%%%%%%%%%%%%%%%%%%%%%%%%%%%%%%%%%
%

%
%%%%%%%%%%%%%%%%%%%%%%%%%%%  FIGURE   %%%%%%%%%%%%%%%%%%%%%%%%%%%%%%
%\newpage

%\newpage
 
%\newpage          

\end{document}